\documentclass[twocolumn,journal,11pt]{IEEEtran}
\usepackage{graphicx}
\usepackage{bm,amsmath}
\usepackage{float}
\usepackage{cite}
\usepackage{epsfig}
\usepackage{caption,subfig}
\begin{document}

\title{Fundamental Limits of Video Coding: A Closed-form Characterization of Rate Distortion Region from First Principles}

\author{

\IEEEauthorblockN {Kamesh Namuduri and Gayatri Mehta\\}
\IEEEauthorblockA{Department of Electrical Engineering \\
University of North Texas}
}

\linespread{1}
\maketitle

\linespread{1}
\begin{abstract}

Classical motion-compensated video coding methods have been standardized by MPEG over the years and video codecs have become integral parts of media entertainment applications. Despite the ubiquitous use of video coding techniques, it is interesting to note that a closed form rate-distortion characterization for video coding is not available in the literature. In this paper, we develop a simple, yet, fundamental characterization of  rate-distortion region in video coding based on information-theoretic first principles.  The concept of conditional motion estimation is used to derive the closed-form expression for rate-distortion region without losing its generality. Conditional motion estimation offers an elegant means to analyze the rate-distortion trade-offs and demonstrates the viability of achieving the bounds derived. The concept involves classifying image regions into active and inactive based on the amount of motion activity. By appropriately modeling the residuals corresponding to active and inactive regions, a closed form expression for rate-distortion function is derived in terms of motion activity and spatio-temporal correlation that commonly exist in video content. Experiments on real video clips using H.264 codec are presented to demonstrate the practicality and validity of the proposed rate-distortion analysis. \end{abstract}


\IEEEpeerreviewmaketitle
\linespread{1}

\section{Introduction}

\begin{figure}[h] \begin{center} \includegraphics[width=5in]{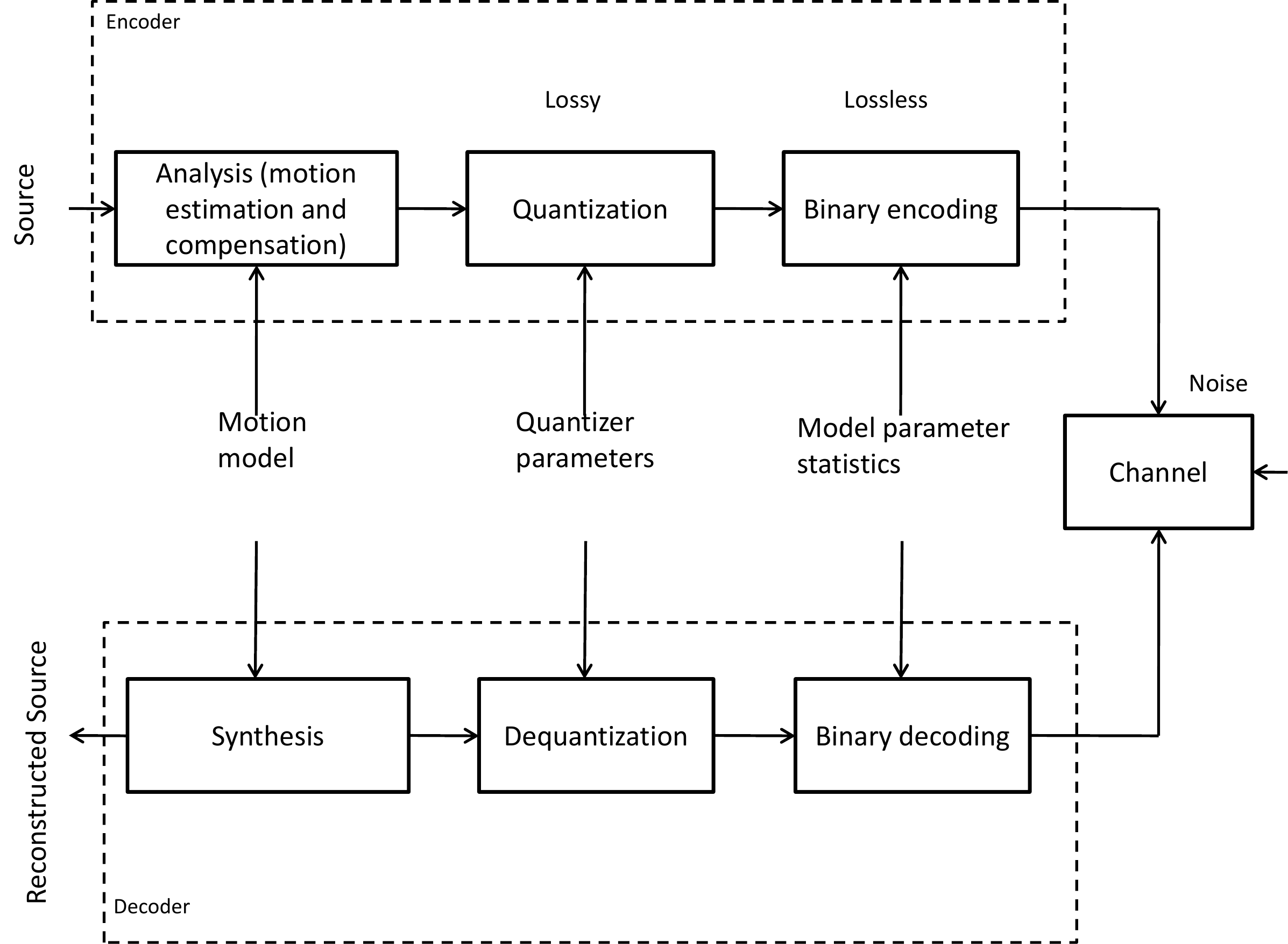}
\caption{A classical video encoding system}
\label{fig:Coding}
 \end{center}	
\end{figure}

How much compression is possible on a video clip? The analysis presented in this paper answers this question by deriving rate-distortion bounds for video coding  from first principles. A classical video coding system consists of three main components \cite{Wang2002}(Chapter 8): (1) video analysis which includes motion estimation and compensation, (2) quantization, and (3) binary encoding  as depicted in Fig. \ref{fig:Coding}. The effectiveness of a video coding technique depends on the
rate distortion tradeoffs that it offers. Often, an
application such as a video streaming service might dictate the limit on maximum acceptable
distortion. This limit, in turn, places a bound on the minimum
amount of rate required to encode video. On the other hand, a communication technology may put a limit on the maximum data transfer rate that the system can handle. This limit, in turn, places a bound on the resulting video quality. Therefore, there is a
need to investigate the tradeoffs between the bit-rate (R) required to encode video and the resulting distortion (D) in the reconstructed video. 
Rate-distortion (R-D) analysis deals with lossy video coding and it establishes a relationship between the two parameters by means of a \emph{Rate-distortion} $R(D)$ function \cite{Shannon1960,Cover2004,Berger1971}. Since R-D analysis is based on information-theoretic concepts, it places absolute bounds on achievable rates and thus derives its significance.  
 
\subsection{Motivation}

Motion estimation and compensation process is one strategy to remove the temporal correlation that naturally exists in video. This is accomplished in practical video coding standards  such as MPEG-4 and H.264 by means of block-based motion estimation methods.  
While video coding is an integral part in  every multimedia application that exists today, it is interesting to note that a closed form rate-distortion characterization for video coding is not available in the literature.
Rate-distortion optimization (RDO) \cite{ortega1998,Shoham 1988,Sullivan1998}, which is based on Lagrangian formulation comes close to the R-D analysis presented in this paper. While RDO is mathematically elegant and has been widely and effectively used in many practical coders including H.264, it doesn't lend itself to closed form characterization of R-D tradeoffs. The proposed R-D analysis, on the other hand, is based on information-theoretic concepts and thus establishes absolute bounds on R-D tradeoffs. These bounds are based on measurable and quantifiable parameters such as motion activity and correlation  in video. Hence, they are of significant practical value.

\subsection{Main Contributions}

In this paper, we develop a theoretical basis to investigate the R-D tradeoffs in video coding. We characterize the R-D region using the concept of conditional motion estimation. 
In this concept, motion estimation and compensation are performed only for active regions selected based on a motion activity criterion \cite{Rath2003,Ramya2005}. Based on this concept, we devise a strategy to balance the bit-budget (rate) against the video quality. We derive a closed-form $R(D)$ relationship for video coding from the first principles and validate this relationship by experimenting with several video clips.  The main contributions of this paper are outlined below.

\begin{itemize}

\item First, a framework for conditional motion estimation and compensation that facilitates the derivation of R-D formulation, is presented.
In this framework, each frame is divided into blocks of standard size. The blocks are
classified into active and inactive categories based on the magnitude of intensity change between consecutive frames. The residuals corresponding to active blocks
are represented by  \emph{displaced frame differences} (DFDs).  The DFDs are quantized and then encoded along with motion vectors.
The residuals corresponding to inactive
blocks are represented by \emph{frame difference} (FDs).
FDs are quantized and then encoded without motion vectors. We demonstrate that
this process offers an excellent  means to analyze the R-D tradeoffs  in video coding.

\item Second, the R-D function associated with 
motion estimation and compensation is derived from first principles. Using conditional motion estimation doesn't reduce the generality of the R-D function derived. On the other hand, conditional motion estimation provides a means to derive the R-D function.
If motion estimation is  perfect,  then DFDs corresponding to active blocks will be uncorrelated \cite{Wang2002} and can be modeled as white Gaussian process. While perfect estimation of motion is not practical, it is meaningful and leads to the derivation of an upper bound for R-D tradeoff. 
On the other hand, the residuals corresponding to inactive pixels, i.e., FDs, exhibit spatio-temporal correlation and hence are modeled using Gauss-Markov process. For simplification purposes, we use first-order Gauss-Markov model parametrized by the correlation coefficient  $\rho_I$, where the subscript $I$ indicates inactive blocks. Modeling of FDs using Gauss-Markov process leads to the derivation of a lower-bound for R-D tradeoff. Different videos exhibit different levels of motion activity, which we quantified using the parameter $\lambda_M$ that varies between 0 to 1.  We derive a closed form expression for the R-D region in terms of $\rho_I$ and $\lambda_M$.
This closed form expression is applicable to all classical motion compensated video coding schemes that exist today and is given by:

 \begin{equation}
R=\log_2\left[  \left(\frac{\sigma_A^2}{D_A}\right)^\frac{\lambda_M}{2} \left(\frac{\left(1-\rho_I^2\right)\sigma_I^2}{D_I}\right)^\frac{\left(1- \lambda_M\right)}{2}\right]
   \label{RD_overall2}
  \end{equation}
 
where ($\sigma_A^2,D_A$) and ($\sigma_I^2,D_I$) are  the variance and distortion pairs associated with the uncorrelated and correlated residual streams resulting from the motion estimation and compensation process. Details of the derivation are given in Section IV.

\item
Third, the proposed R-D formulation is validated by experiments on several video clips. In particular, we used H.264 codec to compress and reconstruct videos with different levels of motion activity. 
Results demonstrating the validity of the proposed model are presented on two specific video clips: one with low motion activity and one with high motion activity.

\end{itemize}


\subsection{Organization}

This paper is organized as follows. Section \ref{background} presents a brief
literature survey in R-D theory and its applications. It also discusses the related work
in conditional motion estimation. Section \ref{conditional} explains the
conditional motion estimation concept. The R-D analysis for  motion estimation based video encoding is discussed in section \ref{RDAnalysis}. Section \ref{exps} outlines experimental results, discussions, summary, and conclusions.

\section{Related Work}
\label{background}

There is an extensive literature on R-D theory and R-D optimization methods for video encoding.  Below, we provide a brief summary of research progress applicable to the issues discussed in this paper.

\subsection{Rate-Distortion Analysis}

The foundation for R-D theory was first formulated by
Shannon in \cite{Shannon1960}. A survey of early contributions to  R-D theory are presented in \cite{Andrews1971}. R-D analysis for scalable video coding is presented in \cite{Cook2006,Zhang2010}. Closed form $R(D)$ expressions have been derived in the
literature for different types of sources, not necessarily for video. For example, the $R(D)$ function for an independent and identically distributed (i.i.d.) Gaussian source is given by \cite{Cover2004} (Ch. 13),
\begin{eqnarray}
R(D) & =&
\frac{1}{2}\log_{2}{\frac{\sigma^{2}}{D}},\label{eq:D-dfd00}
 \end{eqnarray}
\noindent where $\sigma^{2}$ is the variance of the source. If the source samples are correlated as in the case of a typical video, this formulation is not applicable. A R-D function for any source that can be modeled as an
$N^{th}$ order Gauss-Markov process is derived in
\cite{Ghoneimy1993}.

\subsection{Rate-Distortion Analysis for Video Coding}

One way to improve the efficiency of video coding is by exploiting the spatial and temporal correlations that exist in the video through the use of vector coding. In practice, vector coding is implemented using block-based methods. Deriving an
$R(D)$ function corresponding to block-based video coding is much more complex compared to that of scalar coding \cite{Chou1989,Bunin1969}.
While video encoders always make use of block based strategies, they are not easily amenable for R-D analysis.

An alternate method for deriving the R-D relation for video encoding is by modeling the process of  motion estimation and compensation. One of the most widely used and effective rate distortion optimization (RDO) technique in video coder control is Lagrangian formulation \cite{Shoham1988,Sullivan1998,Chou1989}, which is discussed below.

\subsection{Rate-Distortion Optimization}

The Lagrangian formulation of the rate distortion optimization problem is given by,

\begin{equation}
min\{J\}, ~where ~J = D + \lambda_L R,
\end{equation}
where the Lagrangian rate-distortion functional, $J$, is minimized for a particular value of the Lagrangian multiplier $\lambda_L$ \cite{ortega1998,Shoham 1988,Sullivan1998}. 

In practical coders such as H.264, RDO is used in both motion
estimation to find a motion vector and the subsequent mode decision to decide a suitable mode to encode
the residual data. While the applicability of RDO formulation is demonstrated by practical coders, it doesn't lend itself to a closed form characterization of the R-D region which is important for theoretical analysis. This is the focus of our research work.

\subsection{Applications Beyond Coding}
Applications of R-D analysis extend beyond coding. A study of the R-D function and its applicability in designing a practical communication system for video sources with bounded performance is discussed in \cite{Davisson1972}. Numerous
other applications of R-D theory beyond coding, communications and signal processing are extensively
discussed in \cite{Gallager1968}. In many applications, a basic R-D problem is formulated and solved using techniques such as Lagrangian. In pattern classification \cite{Pearl1976}, for example, features belonging to
different classes are assumed as outputs of a source and an
equivalent data compression problem is designed. The
\emph{R}(\emph{D}) function for such a data compression problem explains
the tradeoffs between the number of features selected and the
resulting error in classification.

\subsection{Paradigm Shift in Video Encoding}

Conventional video coding techniques perform motion estimation on
the sender side. Motion in video is represented using different
methods including pixel-based representation, region-based
representation, block based representation, and mesh-based
representation etc \cite{Wang2002}. Motion is estimated using a
criterion such as DFD or Bayesian. A tutorial on estimating two
dimensional motion is presented in \cite{Stiller1999}.

The complexity of an encoder increases as the complexity of the
motion estimation method increases. An encoder of this kind is not
suitable to be used in resource constrained application such as
wireless sensor networks. A better way of encoding in resource
constrained situations is \emph{distributed source coding} which
is built on Slepian - Wolf coding, \cite{Slepian1973}, Wyner-Ziv
coding \cite{Wyner1974} and channel coding principles. One of the
coding techniques built on distributed source coding principles, known as
PRISM, is described in \cite{Puri2007}. The principles of
distributed source coding \cite{Slepian1973} are extended to
lossy-compression in \cite{Wyner1976}. The R-D analysis for Wyner-Ziv video coding has been proposed in \cite{ZLi2007}.

Rate-distortion in distributed systems has applicability in video surveillance networks. A rate-distortion function for distributed source (video) coding with
$L+1$ correlated memoryless Gaussian sources in which L sources
are assumed to provide partial side information at the decoder
side to construct the $L+1^{th}$ source is proposed in \cite{Oohama2005}.

\section{Conditional Motion Estimation}
\label{conditional}

In this section, we briefly outline the process of conditional
motion estimation. Conditional motion estimation process begins with subdivision of
the image frames into
blocks of equal size, and then, classification of  these blocks into active
or inactive classes. Activity is determined based on the difference in intensities corresponding to two consecutive frames. Then, block-based motion estimation is performed for only active blocks.

\subsection{Active and Inactive Blocks}
\begin{figure}[h]
    \centerline{\hbox{ \hspace{0.0in}
		\includegraphics[width=1.5in]{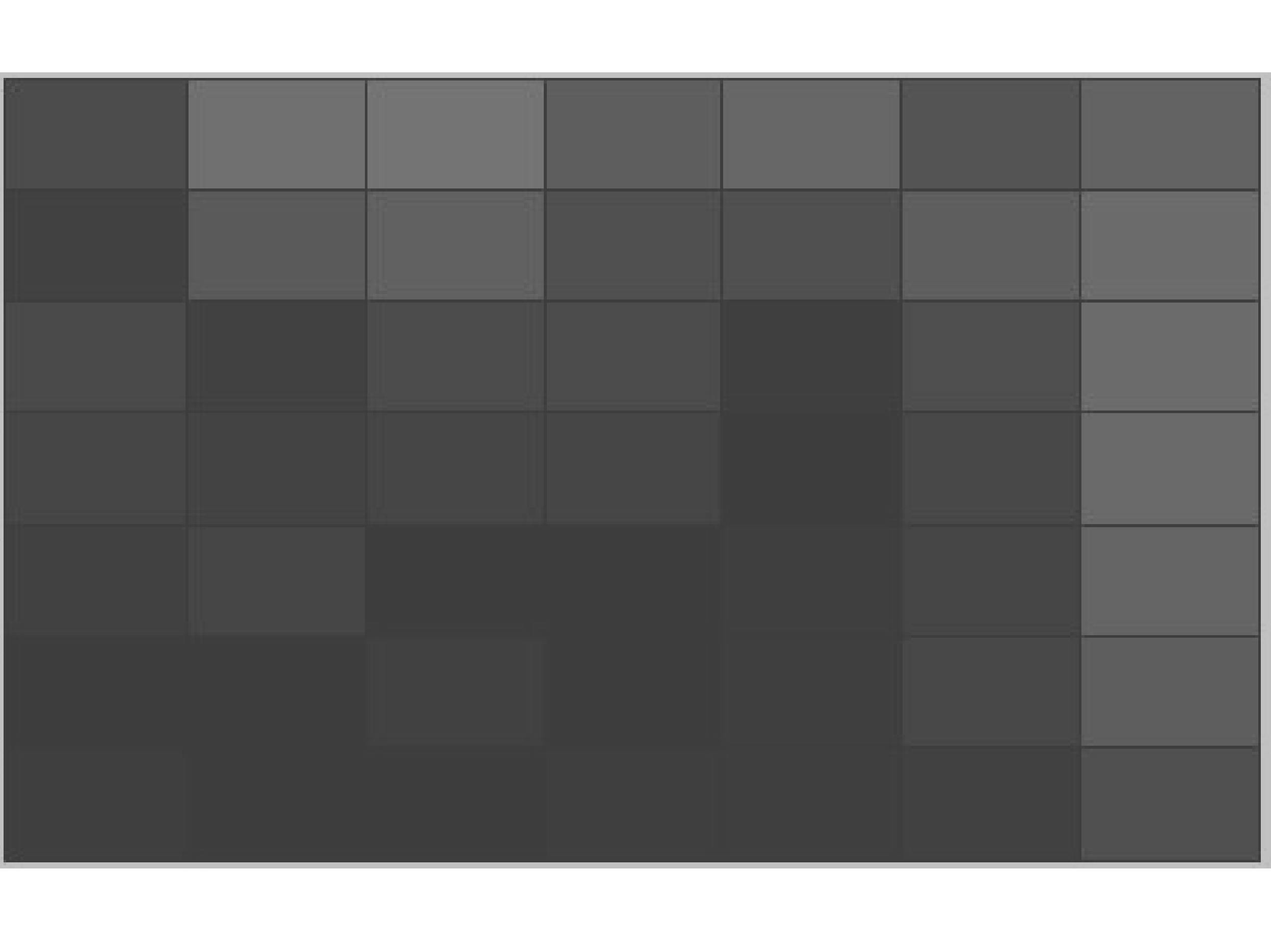}
		\hspace{0.20in}
		\includegraphics[width= 1.5 in]{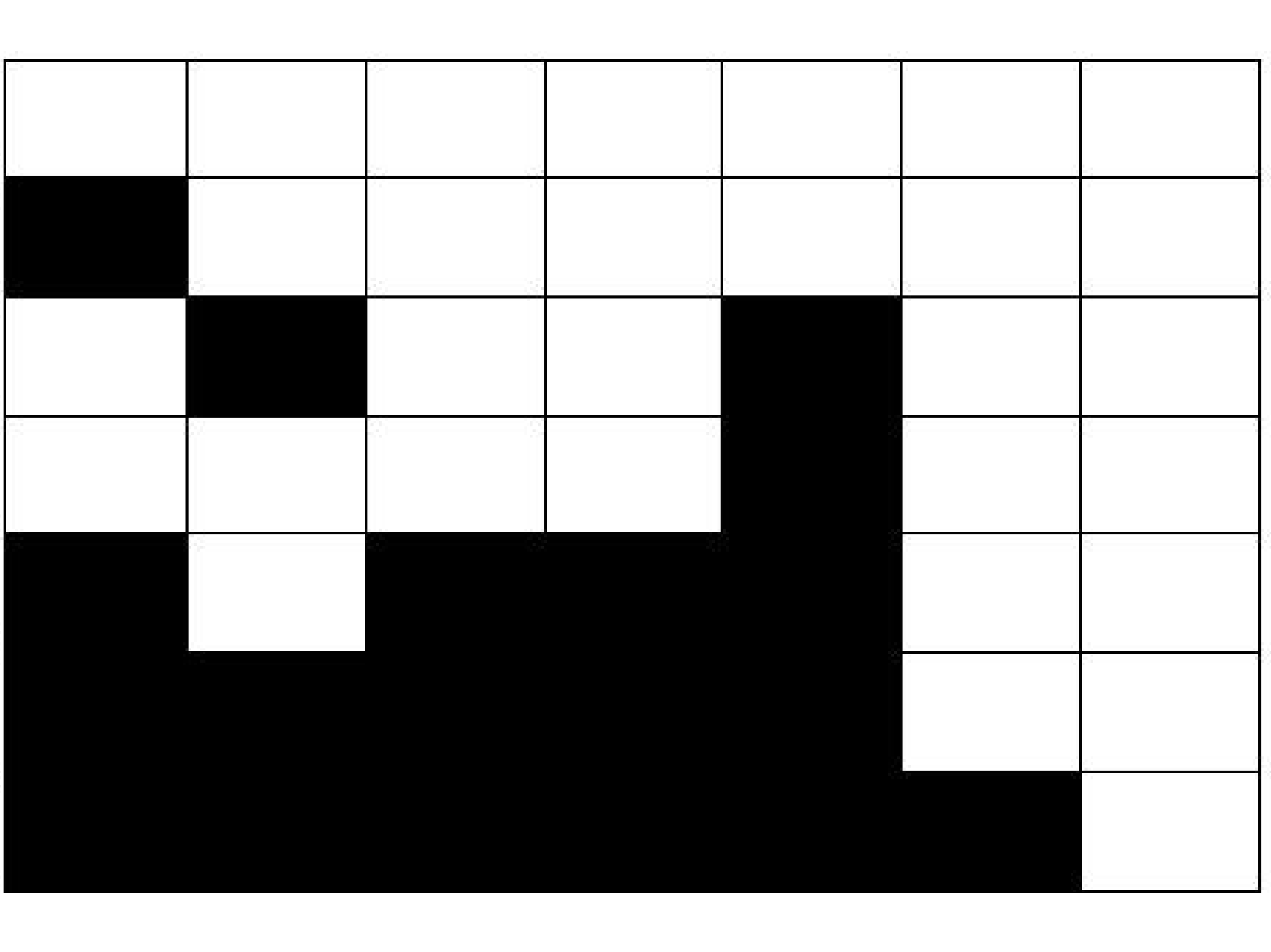}
 }
  }
    \hbox{\hspace{2.5 in} (a) \hspace{1.65in} (b)} 
 \caption{\small Figure (a) depicts a block in a difference
image and figure (b) depicts the active pixels in that block. The
gray pixels in (a) indicate the intensity values and the dark and
bright pixels in (b) indicate the pixels with intensities above
$T_{g}$ and below $T_{g}$ respectively.} \label{fig:pixels}
\end{figure}

The classification of the blocks into active and
inactive blocks is based on two thresholds, one at pixel level
$(T_{g})$ and one at block level $(T_{p})$ \cite{Rath2003}. If a
value in the difference image is greater than the threshold
$T_{g}$, then that pixel is classified as an active pixel,
otherwise, it is classified as an inactive pixel \cite{Rath1999}.
The two thresholds need to be chosen adaptively based on the spatial and temporal correlations present in the video and the desirable level of R-D tradeoff. In our previous work, we developed an online training strategy to adaptively select the thresholds using Bayesian criterion in \cite{Ramya2005,Sridevi2004}.
Fig. \ref{fig:pixels} depicts a
block in a sample difference image and the active and
inactive pixels within the block. The
number of active pixels in every block is counted, and if this
count is greater than $T_{p}$, it is classified as an active
block, else, it is classified as an inactive block.

The selection of the two thresholds $T_{g}$ and $T_{p}$ is an
important task in conditional motion estimation process. The
selection of $T_{g}$ is crucial since it directly decides if a
pixel is active or not. It is known that in a frame there exits a
correlation between intensities of adjacent groups of pixels.

\begin{figure}[h]
	\centering
		\includegraphics[width =3.2in]{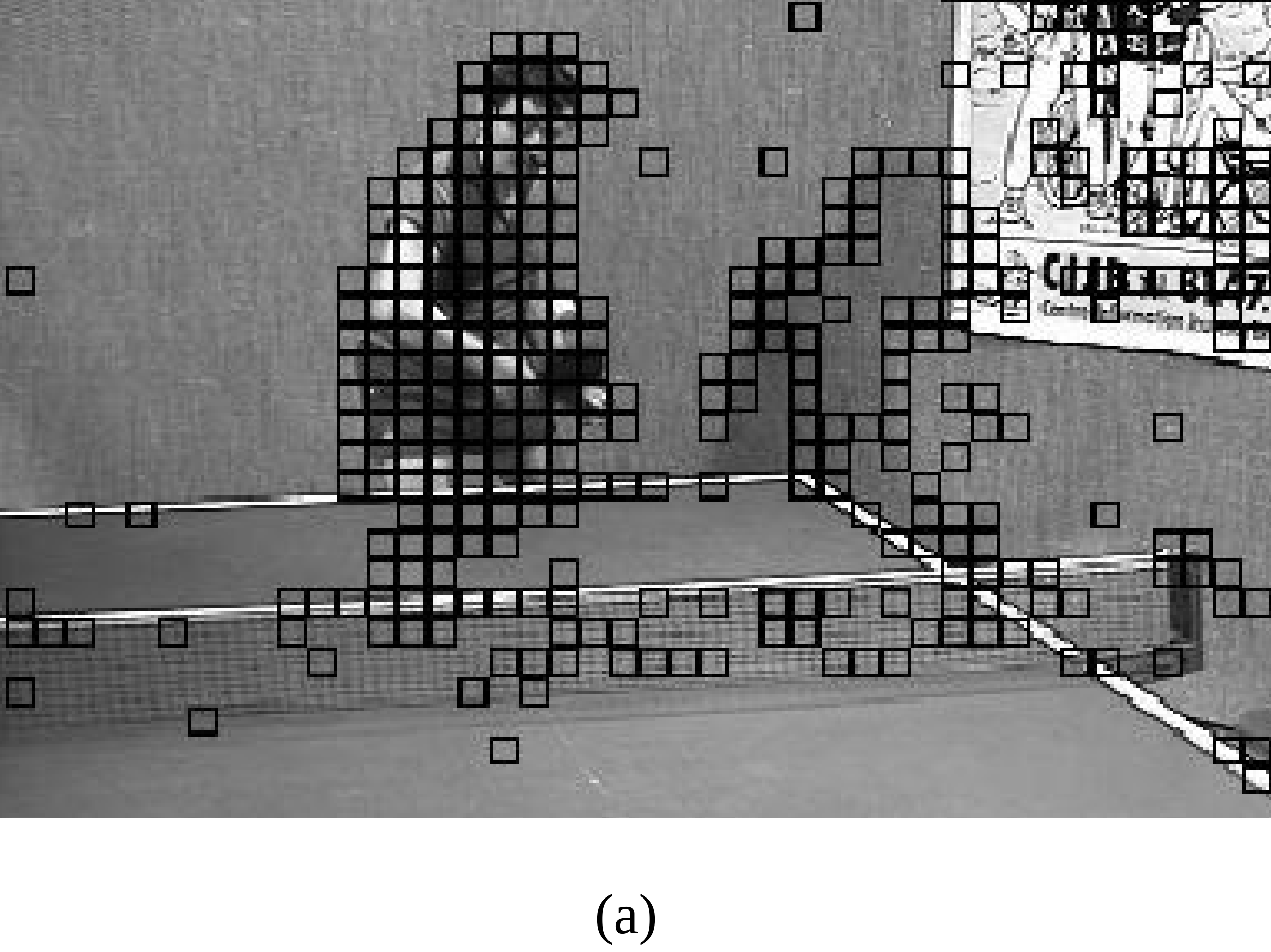}
    \vspace{1pt}
    \includegraphics[width =3.2in]{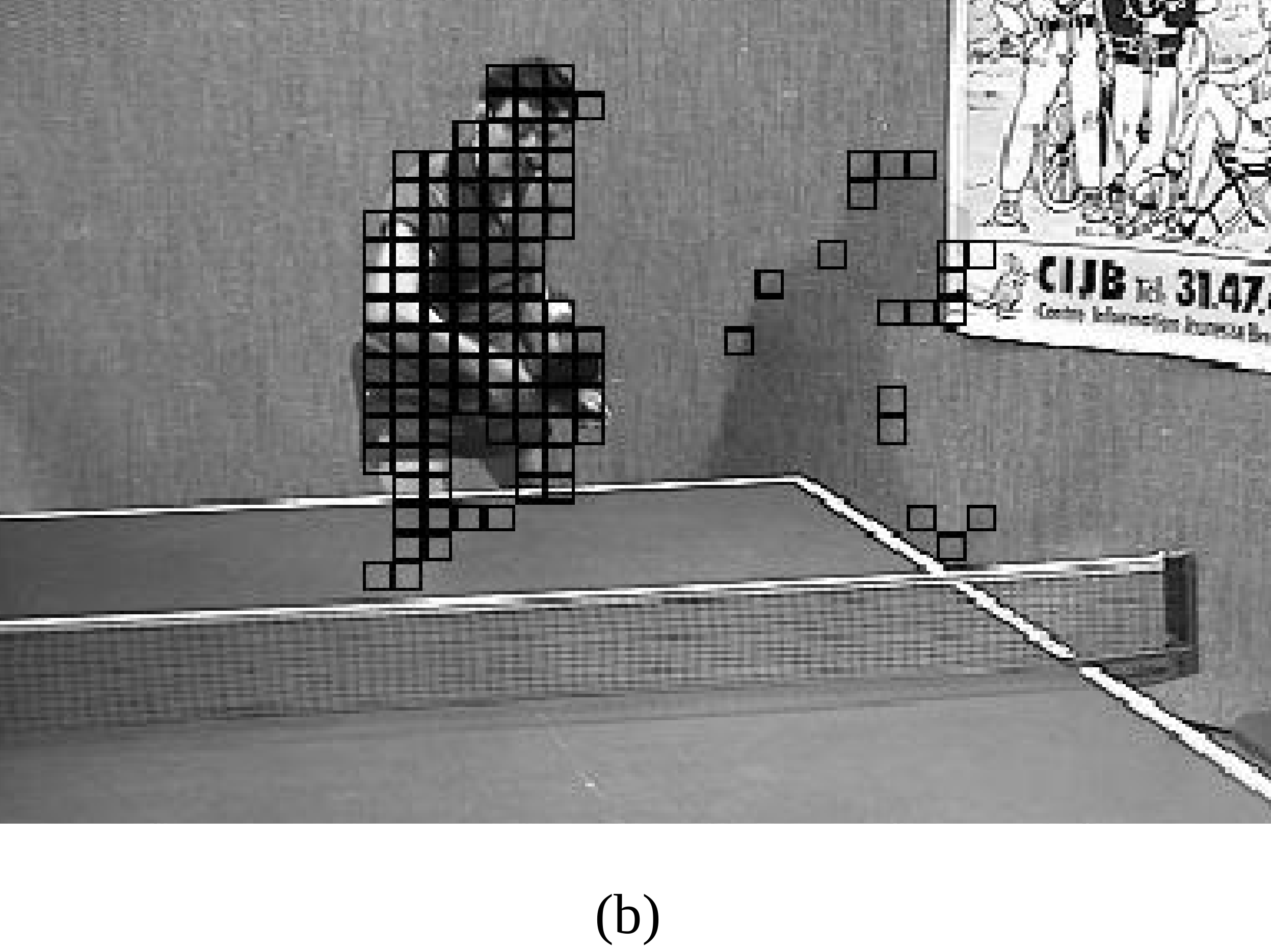}
    \vspace{-0.1in}
    \caption{\small The figure illustrates the impact of $T_{p}$ on the
number of active blocks in a frame. (a) smaller value of $T_{p}$,
say 8, results in a large number of active blocks and (b) larger
value of $T_{p}$, say 32, results in a small number of active
blocks. }\label{fig:Tp}
\end{figure}

The selection of $T_{p}$ also significantly impacts the
performance of the proposed method. For example, if $T_{p}$ is
increased, then the number of active blocks decreases, resulting
in low bit rate and high distortion. On the other hand, if $T_{p}$
is decreased, then the number of active blocks increases,
resulting in high bit rate and low distortion. In order to
demonstrate this, an image is selected from a video sequence and
the difference image (i.e., the difference between the current
frame and previous frame) is found. The active blocks found in
this frame using the difference image are displayed for two values
of $T_{p}$ in Fig. \ref{fig:Tp}. It can be observed that when
$T_{p}$ is small, the number of active blocks is large and
vice-versa. In our approach, $T_{p}$ is kept constant for all the
blocks in order to simplify the analysis.

\subsection{Difference Image}

Let $F_{1}(\overline{\emph{x}})$ be the anchor frame and
$F_{2}(\overline{\emph{x}})$ be the target frame. If
D($\overline{\emph{x}}$) represents a difference image, then,

\begin{eqnarray}
D(\overline{\emph{x}})& =& |F_{2}(\overline{\emph{x}}) -  F_{1} (\overline{\emph{x}})|,\label{eq:E-diff}
\end{eqnarray}
where $\overline{\emph{x}}$ is a vector
representing pixel locations. Every pixel in
$D(\overline{\emph{x}})$ is compared to its corresponding $T_{g}$
and classified as an active pixel if it is greater or inactive if
it is lesser. The number of active pixels in a block are then
counted and if the count is greater than $T_{p}$, then that block is
classified as an active block else it is classified as an inactive
block. Once the active blocks in a target frame are determined,
the next step is to perform block-based motion estimation for all
those active blocks. We assume that the anchor frame is already
available at the decoding side and encode the target frame using
conditional motion estimation.

\subsection{Block-based Motion Estimation}

Block-based motion estimation is a motion compensated video
technique used in various video coding standards including $H.26X$
\cite{CCITT1989}, MPEG-X \cite{ISO1993}. A block-based motion estimation technique uses a
block-matching algorithm to test each block in the anchor frame
with every block in the target frame to find the block that
matches the most. The matching criteria is usually the mean
square difference between the blocks compared. In our work, a
fast block-matching search algorithm called \emph{diamond search
algorithm} \cite{SZhu2000} is used to estimate motion vectors for
all blocks in the anchor frame. A motion vector represents the
displacements of a block along \emph{x} and \emph{y} directions.

Let $\overline{\emph{a}}$ be a motion vector whose parameters
\emph{$a_h$} and \emph{$a_v$} represent the horizontal and
vertical displacements that a block in an anchor frame undergoes
to reach its position in the target frame. If every block is
identified by the first pixel in it, then the set of motion
vectors for the frame can be represented as
$\emph{d}(\overline{\emph{x}} ; \overline{\emph{a}})$
\cite{Wang2002}. In conditional motion estimation, we find motion
vectors for active blocks only. The inactive blocks are assumed
not to have moved and are represented with zero motion vectors.
Once the motion vectors are found, the displaced frame difference,
$E(\overline{\emph{x}})$, which is the difference between the target frame and the
motion compensated anchor frame, is generated. This can be written
as,

\begin{eqnarray}
  E(\overline{\emph{x}})&=&F_{2}- C(F_{1}; \emph{d}),\label{eq:E-dfd}
\end{eqnarray}

\noindent where $C(F_{1}; \emph{d})$
is the motion compensated frame constructed from
$F_{1}(\overline{\emph{x}})$ and motion vectors set
$\emph{d}(\overline{\emph{x}} ; \overline{\emph{a}})$. The
displaced frame difference, thus evaluated, is scalar quantized to
obtain $Q_{D}(\overline{\emph{x}})$.

\section{Rate-Distortion Analysis}
\label{RDAnalysis}

A video encoder based on conditional motion estimation effectively transforms the video into an alternate  representation consisting of three different outputs: (1) motion vectors, (2) quantized DFDs corresponding to active blocks, and (3) quantized FDs corresponding to inactive blocks. In many practical video coding, transformations such as Discrete Wavelet Transform  (DWT) and Discrete Cosine Transform (DCT) are employed.  They do not impact the theoretical R-D analysis because of their orthogonality and energy-preserving properties. Hence, they are not taken into account in our analysis. Similarly, while quantization strategies can be taken into account to further fine tune the R-D bounds, they are not considered here to keep the R-D analysis in its simplest and most fundamental form. 

In order to develop R-D analysis for the video coding, one needs to first analyze the R-D tradeoffs offered by these three components individually and later combine them in a meaningful way. In the following subsections, we consider each of these sources, and investigate the corresponding R-D tradeoffs.

\subsection{Motion Activity and Motion Vectors}

Motion activity is a measure of activity level in the video. A sports video clip, for example, will have high activity  whereas a television news clip will have low activity. In clock based encoding methods, motion activity could be measured in terms of proportion of active blocks (say, $\lambda_M \in(0,1)$) in the video. 

Motion estimation and compensation process is an effective strategy to reduce or remove the temporal correlation that usually exists in video. This process transforms the video stream into uncorrelated data stream consisting of motion vectors and residuals. Uncorrelated data stream, in turn, can be encoded using scalar coding as opposed to correlated data stream which requires vector coding. 

The process of motion estimation and compensation leads to a representation that includes MVs and the corresponding DFDs. The bit rate associated with this representation needs to take into account the bit rate required for MVs and the DFDs. There is no distortion associated with MVs. However, quantization of DFDs leads to distortion.

A common model for representing the motion field $\mathit{D}$ is a Gibbs/Markov field  \cite{German1984,Wang2002}. This model is defined by a neighborhood structure called clique. Let $\mathit{C}$ represent the set of cliques; then the probability density function corresponding to Gibbs/Markov field is defined as:

\begin{equation}
P(\it{D} = \textbf{d}) = \frac{1}{Z}exp \left( - \sum_{c \in \it{C}} V_{c}(\textbf{d})\right)
\end{equation}

\noindent where \emph{Z} is a normalization factor. The function $V_c(\textbf{d})$, known as the potential function,  measures the difference between pixels in the same clique:

\begin{equation}
V_c(\textbf{d}) = \sum_{(\textbf{x},\textbf{y}) \in c}   |d(\textbf{x}) - d(\textbf{y})|^2.
\end{equation}

The bit rate allocation for motion vectors, $R_{M}$ is given by,
\begin{eqnarray}
R_{M} &=&\frac{b_{M}}{N_{br}N_{bc}},\label{eq:R-MV}
\end{eqnarray}
\noindent where $b_{M}$ is the number of bits for each motion vector and $N_{br}$ and $N_{bc}$ are the dimensions of the block.

Each active block is represented by $N_{br} \times N_{bc}$ DFDs, and only one motion vector.  Thus, the bit-rate required to represent motion vectors (MVs) is far less compared to the bit-rate  needed to represent FDs and DFDs.  As the block size gets larger, the cost of encoding motion vectors becomes smaller.

Further, the number of motion vectors can be directly computed from the number of active blocks. This relationship between the MVs  and active blocks allows us to directly analyze and assess the R-D  tradeoffs associated with MVs.

It is worth mentioning that motion vectors are also correlated. In our previous work, we developed a motion estimation method that takes into account the correlation among motion vectors \cite{kamesh2004}. In the present analysis, however, the correlation among motion vectors is not considered because it is not as fundamental as the correlation that exists at pixel level. 

\subsection{Rate-distortion Analysis for Displaced Frame Differences}

The DFD image is represented by $E(\overline{\emph{x}})$ and it is computed using (\ref{eq:E-dfd}).  This image consists of residuals obtained after compensating each block in the anchor frame for its motion and subtracting the motion compensated anchor frame from the target frame.

Accurate computation of MVs results in a DFD image consisting of pixels that follow independent and identically  distributed (i.i.d.) samples of Gaussian source \cite{Wang2002} which can be encoded using scalar coding techniques.  In practice, however, the residuals obtained after motion compensation are still correlated. Hence, transforms such as DCT and DWT are applied to the residuals to remove the correlation that still exists after motion compensation. For establishing R-D bounds, however, modelling DFDs as i.i.d. Gaussian source is most appropriate.

The R-D relationship for i.i.d. Gaussian source is
given by \cite{Cover2004}(Ch. 13),
\begin{eqnarray}
R_{A} & =&
\frac{1}{2}\log_{2}{\frac{\sigma_{A}^{2}}{D_{A}}},\label{eq:R-act}
 \end{eqnarray}
\noindent where $\sigma_{A}^{2}$ is the source variance, $D_{A}$ is the distortion resulting from encoding the active pixels.

\subsection{Rate-distortion analysis for Frame Differences}
\begin{figure}[h]
	\centering
		\includegraphics[width=5in]{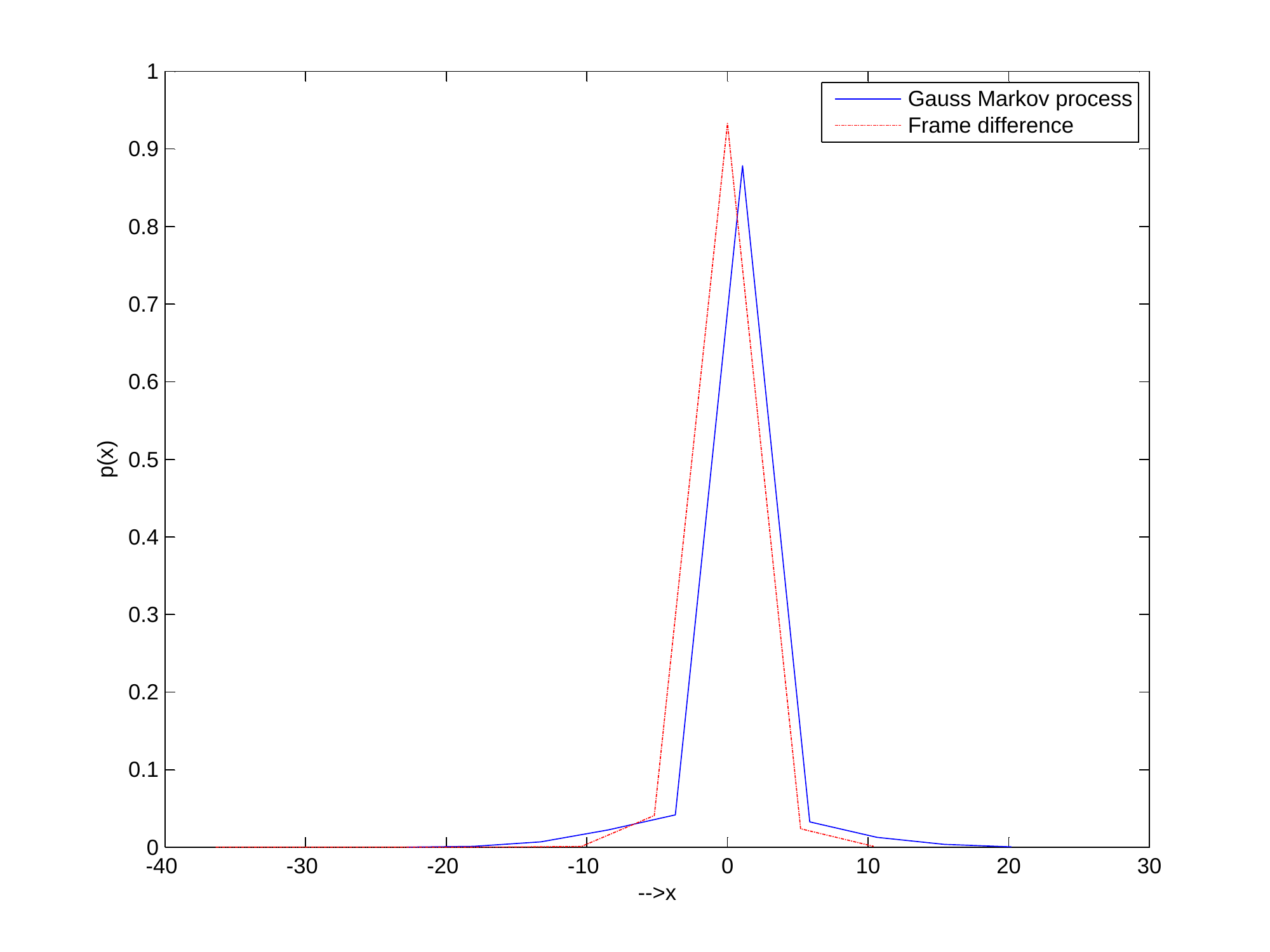}
		\caption{Figure shows the PDF corresponding to a first-order Gauss Markov process and that of frame differences}
	\label{fig:GM-model}
\end{figure}

There exists a strong spatial correlation among pixels in video. In image processing literature, real-world images are best modeled using Gaussian process \cite{Jain1989}, and Gauss-Markov process is an appropriate model for a correlated Gaussian source.  For simplification, we model FD samples using the first order Gauss-Markov process. Several experiments have been carried out to test the suitability of Gauss-Markov process for modeling FDs. The close match between the PDF of Gauss-Markov source and that of the DFDs shown in Fig.  \ref{fig:GM-model},  validates this assumption.

The relationship between the bit rate $R_{I}$ incurred in encoding a first-order Gauss-Markov source  and the resulting distortion $D_{I}$ in its reconstruction is given  by \cite{Bunin1969},

\begin{eqnarray}
 R_{I}&=&\frac{1}{2}\log_{2}{\frac{(1-\rho_{I}^{2})\sigma_{I}^{2}}{D_{I}}},
 \label{eq:R-inact}
\end{eqnarray}

\noindent where $\rho_{I}^{2}$ is the correlation that exists between adjacent samples and $\sigma_{I}^{2}$ is the  variance of the source samples.

\subsection{Overall Rate-distortion Analysis and Characterization of Rate Region}
Based on the analysis presented so far, the considerations for R-D tradeoffs in video encoding can be summarized as follows:
\begin{itemize}

\item{Motion Vectors}:  Motion vectors are computed only for active blocks. The bit-rate required to encode motion vectors can be computed directly from the number of active blocks. Motion vectors have an indirect impact on the resulting distortion. Accurate estimation of motion vectors will significantly reduce the temporal correlation between consecutive video frames at the expense of computational complexity. 

\item{Frame Differences}: Inactive blocks are represented by FDs only. Motion vectors are not needed for inactive blocks.

\item{Displaced Frame Differences}: Active blocks are represented by MVs and DFDs. Scalar coding is sufficient to encode DFDs. Further, even after accounting for motion vectors, active blocks require lesser bit-rate compared to inactive blocks. 

\end{itemize}

The overall rate and distortion for video coding scheme can now be expressed as,
 \begin{eqnarray}
  R&=&\lambda_M(R_{A} + R_{M}) + (1-\lambda_{M})R_{I} \\ \nonumber
  &=& \lambda_M(R_{A}) + (1-\lambda_{M})R_{I} + \lambda_MR_{M} \\ \nonumber
  &=& \frac{\lambda_M}{2}\log_{2}{\frac{\sigma_{A}^{2}}{D_{A}}} + \frac{(1-\lambda_M)}{2}\log_{2}{\frac{(1-\rho_{I}^{2})\sigma_{I}^{2}}{D_{I}}} \\ \nonumber
  &&+\lambda_M(\frac{b_{M}}{N_{br}N_{bc}})
   \label{RD_overall1}
 \end{eqnarray}
As the block size gets larger, the bits allocated for a motion vector gets smaller compared to the bits allocated for DFDs and FDs. Hence, we can express the RD relationship as follows:
  \begin{equation}
R=\log_2\left[  \left(\frac{\sigma_A^2}{D_A}\right)^\frac{\lambda_M}{2} \left(\frac{\left(1-\rho_I^2\right)\sigma_I^2}{D_I}\right)^\frac{\left(1- \lambda_M\right)}{2}\right]
   \label{RD_overall2}
  \end{equation}

This closed form expression indicates that the video encoding rate is primarily a function of motion activity and the statistics of the residual data after motion estimation and compensation. Video encoding methods need to be accurate in estimating motion vectors and in exploiting the correlation that exists after motion compensation in order to provide better rate-distortion tradeoffs.

\subsection{Theoretical Results}

The above R-D analysis reveals that R-D tradeoffs in a video source depend on primarily two aspects:  motion activity and spatio-temporal correlation which are outlined below. They will be followed up in the next section within the experimental analysis.

\begin{figure}[h]
	\centering
		\includegraphics[width =4in]{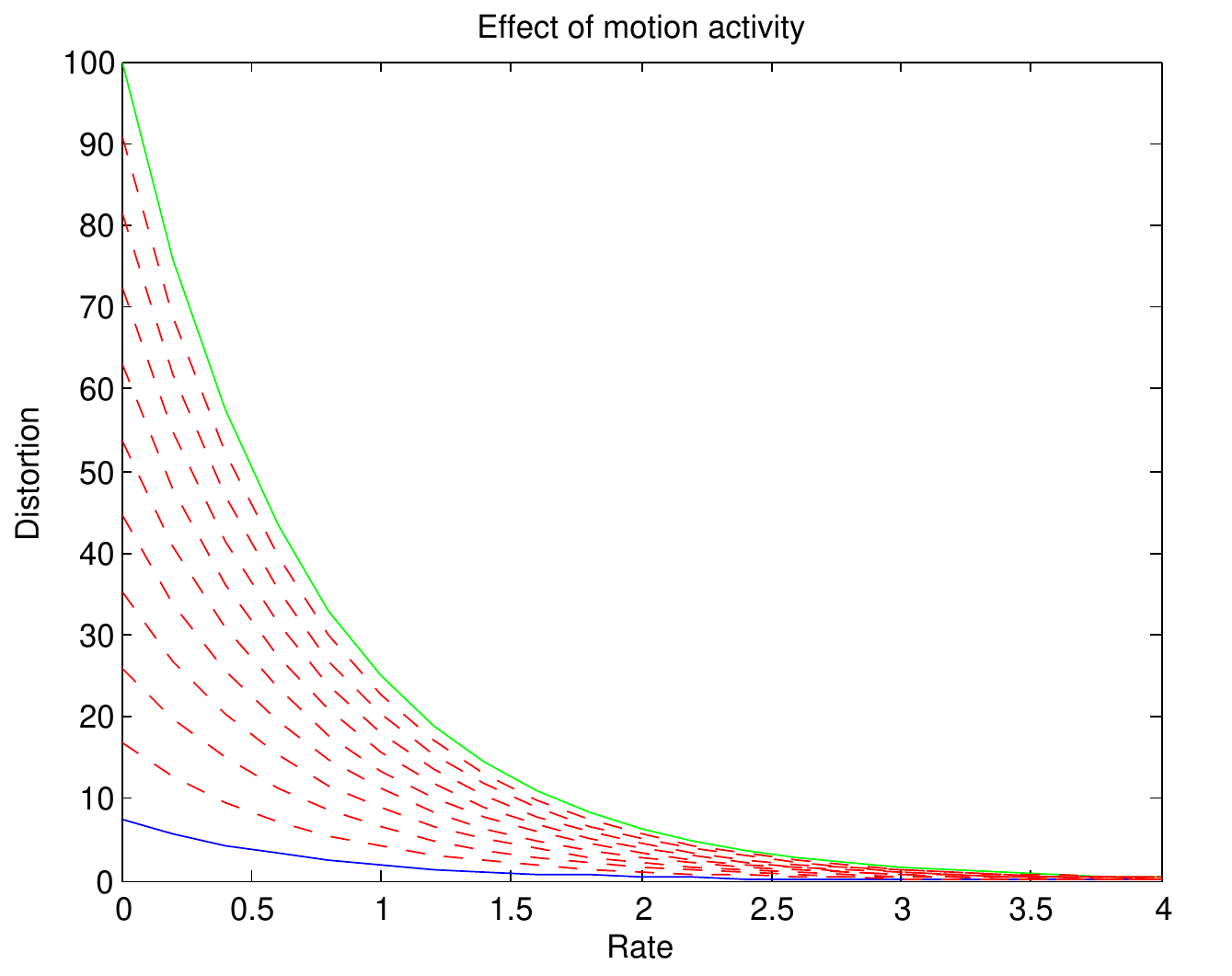}
    \vspace{-0.1in}
    \caption{\small Effect of varying motion activity on R-D: As  $\lambda_M$ increases, the R-D curve moves up}\label{fig:Lambda-Impact}
\end{figure}

\begin{figure}[h]
	\centering
		\includegraphics[width =4in]{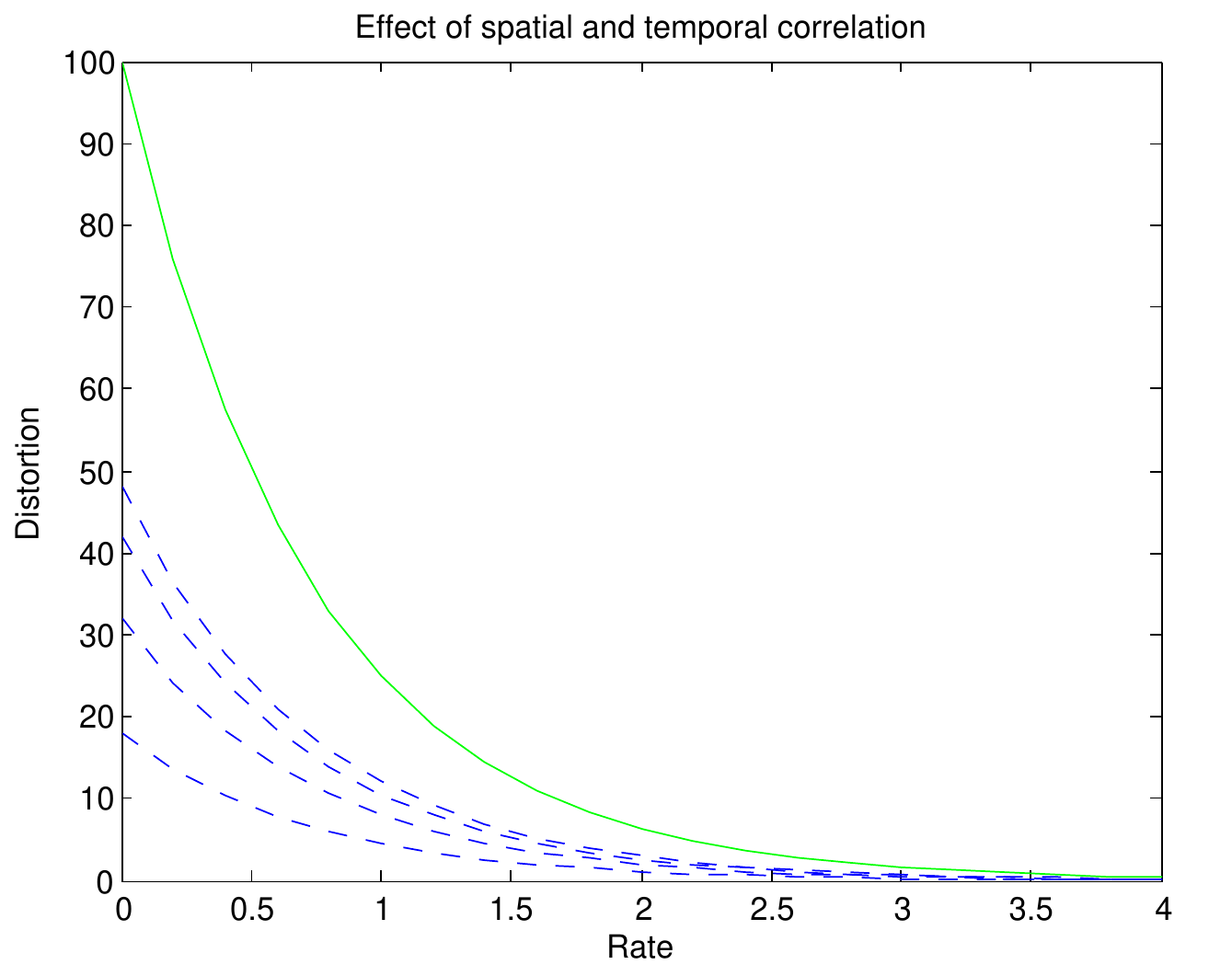}
    \vspace{-0.1in}
    \caption{\small Effect of correlation among the residuals on R-D: As  $\rho_I$ increases, the R-D curve moves down.}\label{fig:Rho-Impact}
\end{figure}

\begin{itemize}

\item{Motion Activity}:  Fig. \ref{fig:Lambda-Impact} shows the variations in R-D curve as a function of  $\lambda_M$. For this experiment, the model parameters are set to the following values: $\sigma^2_{I}$ = 10,  $\sigma^2_{A}$ = 100, and $\rho_I$ = 0.5. 
The lower curve in in Fig. \ref{fig:Lambda-Impact} represents the scenario when the image consists of inactive regions only. This happens for video clips with slow motion. The upper curve in Fig. \ref{fig:Lambda-Impact} represents the scenario when  the image consists of only active regions. This happens for video clips with fast motion.

\item{Spatio-temporal Correlation}: R-D tradeoffs largely depend on spatio-temporal correlation that naturally exists in video sequences. Let the parameter $\rho_I$ represent the  spatio-temporal correlation that exists in the video. 
Fig. \ref{fig:Rho-Impact} shows  the variations in the R-D curve  as a function of the correlation coefficient ($\rho_I$). For this experiment, the model parameters are set to the following values: $\sigma^2_{I}$ = 50,  and $\sigma^2_{A}$ = 100. In this figure, the lower curve represents the scenario with high correlation and the upper curve represents the scenario with less correlation. The plots suggests large correlation leads to large R-D tradeoff. The plots also illustrate that uncorrelated data is difficult to compress.

\end{itemize}

\section{Experiments, Results, and Discussion}
\label{exps}
The proposed R-D analysis has been validated through simulations in MATLAB  and using H.264 codec. The derived closed form bounds for RD bounds are compared with the experimental results obtained from H.264 codec on a wide variety of video clips.


\subsection{Experimental Set up}

The experimental set up consists of H.264 video encoder stimulated on the JM software, which is the official reference software for the H.264/14496-10 AVC profiles. Several video clips of varying resolution (for example, 240 $\times$ 342 and  486 $\times$ 720) in YUV 4:2:0 format with frame rate of 30 frames/sec were encoded into H.264 format and then decoded back to the original file format at different bit rates. The encoder speed ranges from 2.5 Mbits/sec to 30 Mbit/sec. After running the encoder, RD statistics for every encoded frame and cumulative results are collected. Encoder output size in bytes and MSE for Y frame  are recorded for each bit rate. Bit rate (R) for each video is calculated as follows:

\begin{equation}
R = \frac{Encoded~File~Size}{Original~File~size} \times 8~bits/symbol
\end{equation}

The results shown are based on the experiments  on two specific video clips with different levels of motion activity: (1) A \emph{Table Tennis} video clip with low motion activity and (2) a \emph{Football} video clip with high motion activity, each with several number of frames. 


\subsection{Experiments with H.264 Encoder}

\begin{figure}[h] \begin{center}
\subfloat[A target frame from the \emph{Table Tennis} video sequence]
{\label{fig:TableTennisTarget}
\includegraphics[width=0.40\textwidth]{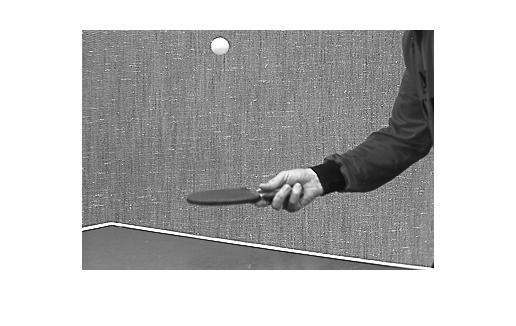}} \hspace{1cm}
\subfloat[A target frame from the \emph{Football} video sequence]
{\label{fig:FootballTarget}
\includegraphics[width=0.40\textwidth]{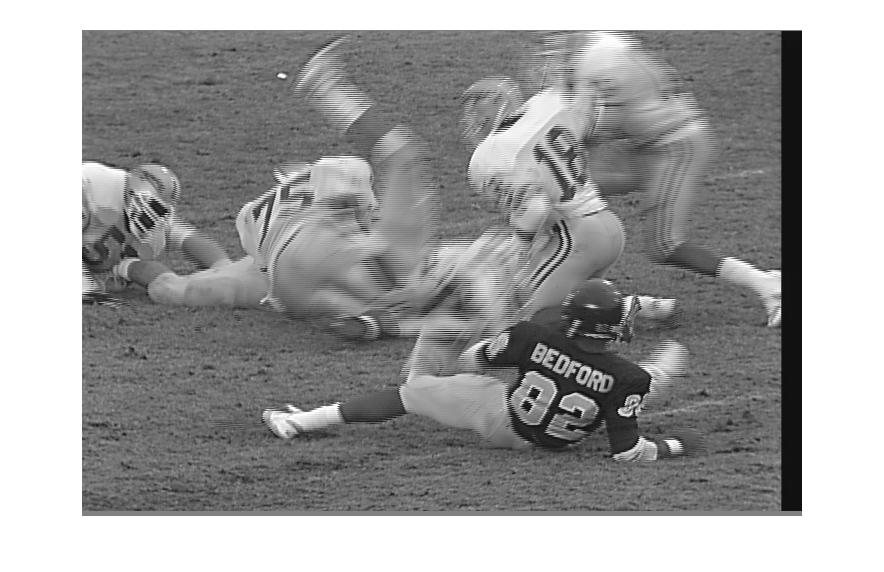}} \\
\subfloat[R-D region for the \emph{Table Tennis} video clip]{\label{fig:TableTennisRD} \includegraphics[width=0.40\textwidth]{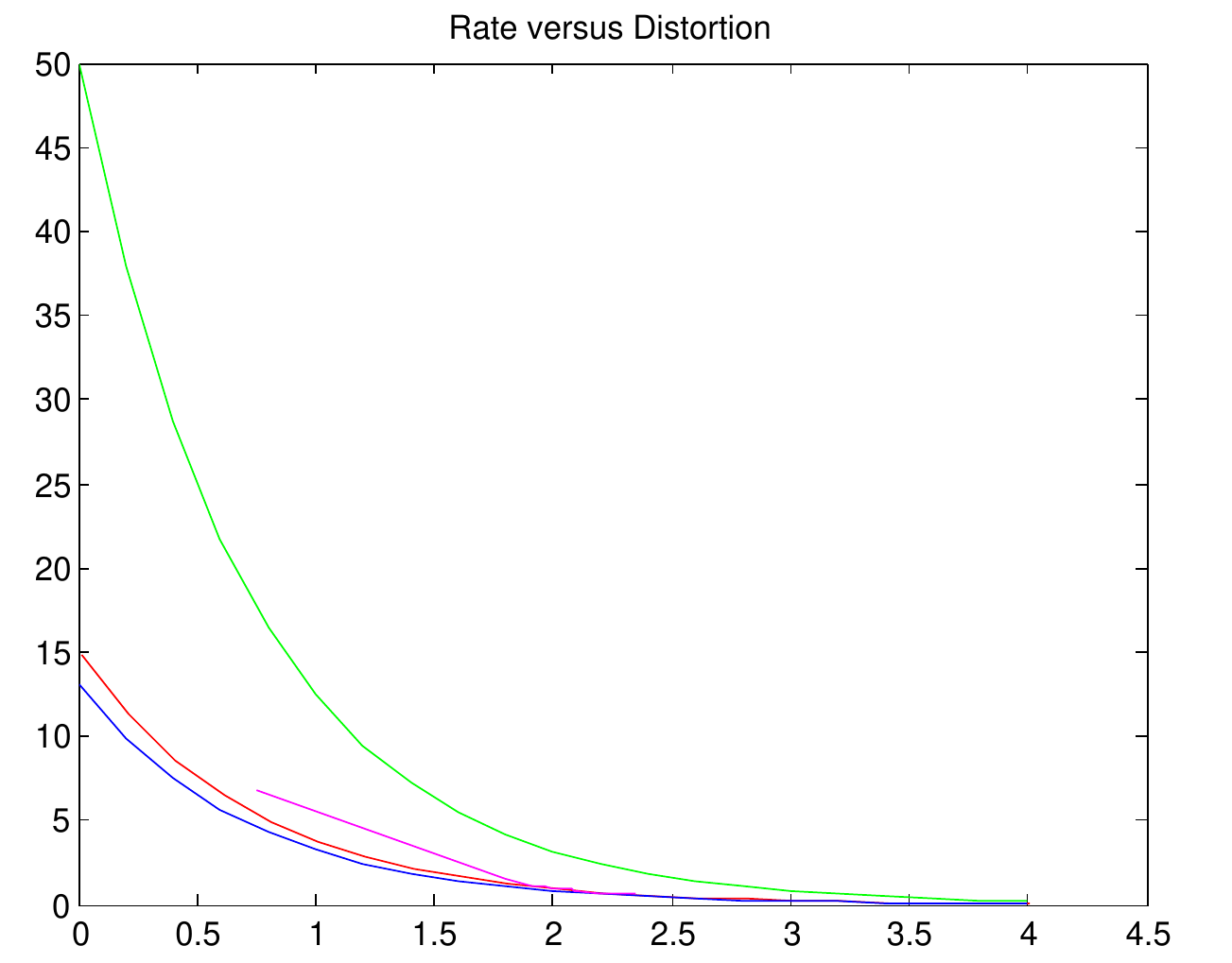}} \hspace{1cm}
\subfloat[R-D region for the \emph{Football} sequence]{\label{fig:FootballRD} \includegraphics[width=0.40\textwidth]{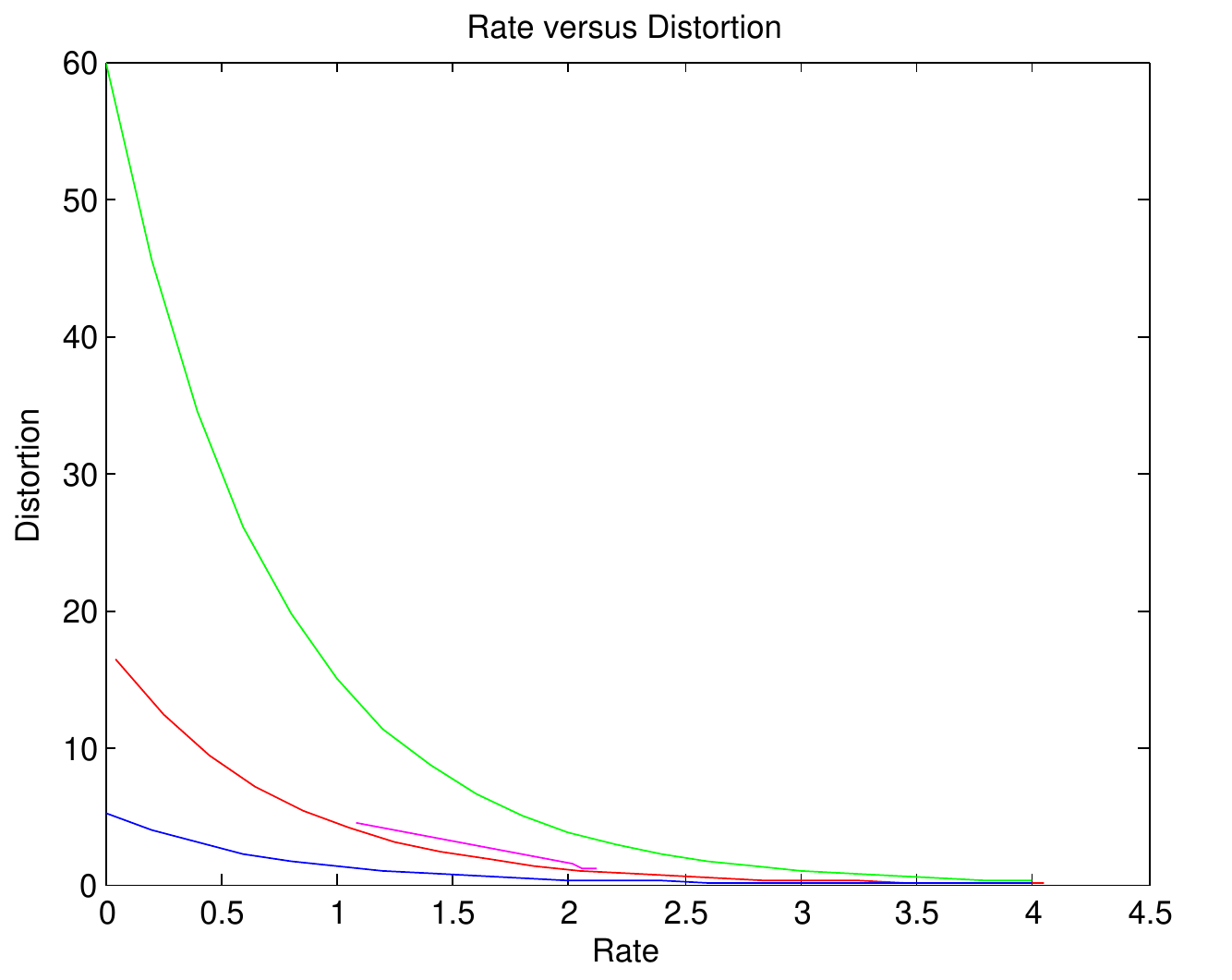}} \\ 

\label{fig:Results}  \caption{Fig (a) and Fig (b) show a target frame in \emph{Table Tennis} and \emph{Football} video clips which are used in experiments. Fig (c) and Fig (d) show the theoretical and practical R-D curves for table tennis and football video clips respectively.}
\end{center}
\end{figure}

Figs \ref{fig:TableTennisTarget} shows a target frame from a \emph{Table Tennis} video clip. The theoretical and experimental R-D results for this video clip are plotted in Figure \ref{fig:TableTennisRD}. The model parameters for this video clip are set to the following values:   $\sigma^2_{I}$ = 20,  $\sigma^2_{A}$ = 50,  $\rho_I$ = 0.59 and $\lambda$ = 0.05.
The plot shown in Magenta color corresponds to the R-D result obtained from the H.264 codec.

Figs \ref{fig:FootballTarget} shows a target frame from a \emph{Table Tennis} video clip. The theoretical and experimental R-D results for this video clip are plotted in Figure \ref{fig:FootballRD}. The model parameters for this video clip are set to the following values:   $\sigma^2_{I}$ = 10,  $\sigma^2_{A}$ = 60,  $\rho_I$ = 0.69 and $\lambda$ = 0.20.

\subsection{Discussion}
The results shown in Fig. \ref{fig:TableTennisRD} and Fig. \ref{fig:FootballRD} demonstrate the following important aspects of the proposed RD model.
\begin{itemize}

\item The top and bottom plots in each figure provide  information-theoretic bounds derived based on the PDFs associated with DFDs, and FDs. The first plot in the middle of each  figure (in Magenta)  represents the R-D results obtained from the H.264 codec. The second plot in the middle of each figure represents the expected R-D results based on the proposed model. The closeness of the theoretical and practical R-D plots demonstrate the validity of the proposed R-D analysis.  
      
\item The region between the two theoretical R-D curves can be characterized as the R-D region for classical motion-estimation based video coding techniques.  This R-D region is dependent on spatial correlation described by ($\rho_I$) and motion activity ($\lambda_M$). 

\item While large spatial correlation makes the R-D curve go down, large motion activity makes the R-D curve go up. This is demonstrated by the experimental plots corresponding to the two video clips.  The second (football) video clip has more motion activity compared to the first video clip, resulting in higher rate as well as larger distortion compared to the first video clip.
\end{itemize}

\subsection{Summary and Conclusions}

In this paper, we characterized the R-D region in video coding using the concept of  conditional motion estimation. Through a practical implementation, we demonstrated the validity of the proposed R-D analysis. Our work can be extended in the following ways.

\begin{itemize}

\item While a typical FD image follows a first-order Gauss-Markov process, it is possible that higher-order Gauss-Markov process can be used to model FD image. This may lead to tighter bounds for the R-D region.
     
\item Information-theoretic R-D analysis doesn't take into account the implementation overheads such as block-based representation, and quantization. Thus, the analysis presented in this paper can be extended and made practical by taking the overheads associated with  implementation.
\end{itemize}


\onecolumn

\end{document}